\documentclass[12pt,nofootinbib]{article}

\usepackage{slashed}
\usepackage{amsmath,amssymb,graphicx,multicol} 
\usepackage{epsf,color}
\usepackage[nosort]{cite}
\usepackage{cancel}
\usepackage{framed}
\usepackage{multirow}

\newcommand{\beq}{\begin{eqnarray}}
\newcommand{\eeq}{\end{eqnarray}}

\newcommand{\centeron}[2]{{\setbox0=\hbox{#1}\setbox1=\hbox{#2}\ifdim

\wd1>\wd0\kern.5\wd1\kern-.5\wd0\fi \copy0

\kern-.5\wd0\kern-.5\wd1\copy1\ifdim\wd0>\wd1
                                       \kern.5\wd0\kern-.5\wd1\fi}}
\newcommand{\ltap}{\>\centeron{\raise.35ex\hbox{$<$}}
                               {\lower.65ex\hbox{$\sim$}}\>}
\newcommand{\gtap}{\>\centeron{\raise.35ex\hbox{$>$}}
                               {\lower.65ex\hbox{$\sim$}}\>}

\newcommand\ZZ{\hbox{\zfont Z\kern-.4emZ}}
\font\zfont = cmss10 

\begin{document}
\begin{titlepage}
\begin{flushright}
\end{flushright}


\begin{center}
{\huge \bf  
Determining Higgs couplings with a \\ \vspace{0.45cm} model-independent analysis of $h \to \gamma \gamma$}
\end{center}
\vskip0.5cm

\renewcommand{\thefootnote}{\fnsymbol{footnote}}
\begin{center}
{\large Aleksandr Azatov, Roberto Contino, Daniele Del Re, \\[0.08cm]
Jamison Galloway, Marco Grassi, Shahram Rahatlou\footnote{email:  aleksandr.azatov@roma1.infn.it, roberto.contino@roma1.infn.it, 
daniele.delre@roma1.infn.it, jamison.galloway@roma1.infn.it, marco.grassi@cern.ch, shahram.rahatlou@roma1.infn.it}
}
\end{center}
\renewcommand{\thefootnote}{\arabic{footnote}}

\begin{center}
{\it Dipartimento di Fisica, Universit\`a di Roma ``La Sapienza'' \\
and INFN Sezione di Roma, I-00185 Rome, Italy} \\
\vspace*{0.1cm}
\end{center}

\vglue 1.0truecm

\begin{abstract}
\noindent 
Discovering a Higgs boson at the LHC will address a major outstanding issue in particle physics but will also raise many new questions. 
A concerted effort to determine the couplings of this new state to other Standard Model fields will be of critical importance.
Precise knowledge of these couplings can serve as a powerful probe of new physics, and will be needed in attempts to accommodate such 
a new boson within specific models. In this paper, we present a method for constraining these couplings in a model-independent way, focusing 
primarily on an exclusive analysis of the $\gamma \gamma$ final state. We demonstrate the discriminating power of fully exclusive analyses, and 
discuss ways in which information can be shared between experimentalists and theorists in order to facilitate collaboration in the task of establishing 
the true origins of any new physics discovered at the LHC.
\end{abstract}

\end{titlepage}

\section{Introduction}
\label{sec:Intro} 

The Higgs boson of the Standard Model (SM) is a special particle in many ways. Its exchange is required to regulate the energy behavior
of the scattering amplitudes involving  longitudinal vector bosons, hence ensuring the perturbative unitarity of the theory in the ultraviolet.
If discovered, it would be the first example of an elementary scalar field, a new form of matter in addition to fermions and the spin-1 carriers of gauge forces.
It would also imply the existence of fundamental forces (its self-interaction and the Yukawa interactions to fermions) not of gauge type.
All these properties, in fact, are strictly related to the possibility for the theory to remain weakly coupled up to extremely large energies, possibly of the order of the
Planck scale. This too, by itself, would be a profoundly new phenomenon in Nature: physics  at the fundamental level
would be described by the same mathematical theory over $\sim 15$ orders of magnitude in energy without any new dynamics appearing at intermediate scales.
To realize this paradigm, the couplings of the Higgs boson must be finely tuned to specific values which depend uniquely on its mass.
Any deviation from these values would  either imply the existence of  additional Higgs bosons that take part in the perturbative unitarization of the scattering 
amplitudes,  like in the case of Supersymmetry, or signal the existence of a new energy threshold at which the theory becomes strongly coupled. In this second case the 
Higgs boson would emerge as a composite state of a new fundamental force~\cite{compositeHiggs}, possibly  of gauge type.

A precise measurement of the Higgs couplings will give the unique opportunity to test the SM paradigm and get  information on the
dynamics behind electroweak symmetry breaking (EWSB). Although  the experimental searches have been so far conducted and, to a large extent, optimized 
in the framework of specific models (\textit{e.g.} the Standard Model or the Minimal Supersymmetric Standard Model), the best strategy to investigate the nature
of the Higgs boson and to report the experimental results is by adopting a model-independent bottom-up approach. The most general description is based
on the  formalism of chiral Lagrangians, supplemented by a few minimal assumptions motivated by the experimental information 
at our disposal. The chiral Lagrangian introduced in Ref.~\cite{Contino:2010mh} and extended by Ref.~\cite{Azatov:2012bz} 
(for earlier related work see~\cite{previousCHL}) fully characterizes
the interactions of a light Higgs-like scalar under the following conditions:
\begin{itemize}
\item new physics states, if present, are heavy and their effect at low-energy can be encoded by local operators in the chiral Lagrangian
\item the EWSB dynamics possesses an (at least approximate) custodial symmetry
\item there are no flavor-changing neutral currents mediated at tree-level by the Higgs.
\end{itemize}
The first assumption implies in particular  that there are no new particles to which the Higgs boson can decay. It can be
easily relaxed by including in the  Lagrangian any new light state that should be discovered. The request of a custodial symmetry is strongly
motivated by the absence of corrections to the $\rho$ parameter measured at LEP and implies that the couplings of the Higgs to the $W$ and the $Z$
must be equal. Under these hypotheses, the  interactions of a single Higgs-like scalar are characterized in terms of a set of parameters which describe the 
couplings to the SM fermions and the electroweak gauge bosons plus new contact interactions to a pair gluons  or photons (as for example generated
by loops of heavy scalar of fermionic top partners). Such a parameter space includes the SM as a specific point, and is sufficiently generic to describe 
scenarios where the Higgs-like scalar is not part of an $SU(2)_L$ doublet or is not even related to  EWSB, as in the case of a light 
dilaton~\cite{dilaton}.~\footnote{By restricting to the case where the Higgs scalar is part of an $SU(2)_L$ doublet, so that the electroweak
symmetry is linearly realized at high energies, the low-energy Lagrangian can be expanded in the usual series of operators with increasing dimension.
In the case of a strongly-interacting light Higgs this formulation coincides with the SILH Lagrangian of Ref.~\cite{Giudice:2007fh}.}

In this paper we demonstrate how exclusive as opposed to inclusive analyses are much more powerful in determining
the Higgs couplings in a model-independent approach. We will do so by focusing on the $h\to \gamma\gamma$ channel, which is the most sensible
in the case of a light Higgs boson. 
We will restrict, for simplicity, to the case in which single Higgs interactions can be parametrized in terms of only two independent parameters:
the coupling to two gauge bosons, $a = g_{hVV}/g_{hVV}^{SM}$,
and the coupling to two fermions $c =g_{h\psi\psi}/g_{h\psi\psi}^{SM}$. New contact interactions mediated by heavy new physics will be assumed to be small and 
to have a negligible impact on the Higgs phenomenology. In this simplified framework we make a first attempt to estimate the precision that 
the LHC can reach on $a$ and $c$ with 2012 data. 
Our results should be compared to previous studies on the measurement of the Higgs couplings, which include 
Refs.~\cite{Zeppenfeld:2000td,Conway:2002kk,Belyaev:2002ua,duhrssen,Duhrssen:2004cv,Lafaye:2009vr,Bock:2010nz} and the more
recent Refs.~\cite{Carmi:2012yp,Azatov:2012bz,Espinosa:2012ir,Giardino:2012ww,Rauch:2012wa,Ellis:2012rx}.
Our exercise also illustrates how the experimental results can and should be reported in a model-independent
fashion.

\section{Exclusive Analysis of the $h\rightarrow \gamma \gamma$ channel}

The sensitivity of the search for the Higgs boson is enhanced when events are divided into categories with different
signal-to-background ratios. This division is also helpful to discriminate among 
different Higgs production mechanisms.  In this analysis we exploit this categorization to improve the constraints 
in the $(a,c)$ plane compared to an inclusive analysis. We use the $h\rightarrow \gamma \gamma$ decay 
since it is the most sensitive channel for a low mass Higgs, and choose $m_h = 120\,$GeV as benchmark value for our analysis.
We adopt this value mostly because  efficiencies and event yields are quite often reported  for this choice of $m_h$ in the experimental papers.

We start from the CMS analyses described in Refs.~\cite{Chatrchyan:2012tw, CMS-PAS-HIG-12-002}. 
Three variables are used to divide events based on the kinematic properties of the $\gamma\gamma$ final state and
 the quality of the photon reconstruction. 
The first variable is the transverse momentum of the $\gamma\gamma$ system,  $p_T(\gamma\gamma)$, which identifies kinematic regions 
with smaller  background contamination. It also enhances the sensitivity to vector boson fusion (VBF) and associated 
production (VH) mechanisms,  which typically produce a  Higgs boson with larger $p_T(\gamma\gamma)$ compared to those produced 
through gluon-guon fusion (GGF).
The second variable, called $R_9$, is related to the shape of the energy deposited by the photon candidates in the electromagnetic
calorimeter and helps to separate events with converted photons.
Finally, the third variable is  the minimum pseudorapidity $\eta$ of the two photons. 
Eight inclusive categories are defined according to the following criteria:
 $p_T(\gamma \gamma)>40\,$GeV and $p_T(\gamma \gamma)<40\,$GeV;  large and small $R_9$;  
and whether both photons are in the central (barrel) region ($|\eta|<1.44$) or at least one photon is in the endcap region  ($|\eta|>1.44$). 
Two additional {\it exclusive}  categories are defined based on the presence of extra jets and leptons in the event, in order
to increase the sensitivity to different production mechanisms.
The first exclusive category ($jj$) includes events with two extra high-$p_T$ jets in the forward region in addition to 
the photon candidates, and is thus enriched with Higgs bosons produced via VBF~\cite{Chatrchyan:2012tw}. 
The selection requires the leading (subleading) jet to have a minimum transverse momentum of $30\,$GeV ($20\,$GeV). 
The two selected jets need to be separated in 
pseudorapidity ($|\Delta\eta_{jj}|>3.5$), and to have a large invariant mass ($m_{jj}>350\,$GeV). 
There is also the additional requirement that the difference between the 
average pseudorapidity of the two jets and the pseudorapidity of the diphoton system (\textit{i.e.} the Higgs boson) 
has to be less than 2.5. 
The second exclusive category ($1l$) includes events with at least one extra lepton and is, therefore, more sensitive to 
Higgs candidates produced via associated production with a $W/Z$ boson, which decays leptonically~\cite{CMS-PAS-HIG-12-002}.
The lepton is required to be isolated and to have a transverse momentum larger than $20\,$GeV and a pseudorapidity which satisfies $|\eta_l|<2.4$.

We use the signal efficiencies and backgrounds estimates reported by CMS in~\cite{Chatrchyan:2012tw}
for the eight inclusive and $jj$ categories, and in~\cite{CMS-PAS-HIG-12-002}
for the $1l$ category.
However, the analysis proposed in this paper requires the knowledge of the signal efficiencies and the expected backgrounds
in each category relative to individual production mechanisms. This information is not available in the referenced CMS analyses
and, therefore, has been estimated at the generator level and extrapolated from the published results.

Di-photon events are generated with MADGRAPH~\cite{Alwall:2007st}  interfaced to PYTHIA 8.130~\cite{pythia}
and are used to estimate the fraction of background with $p_T(\gamma\gamma)$ above and below 40 GeV.
This is done separately in each of the four categories defined by $R_9$ and the photon pseudorapidity.
We assume this fraction to be the same also for the reducible background with at least one fake photon.
This is a reasonable approximation since the reducible background is about 30\% of the total.
For the exclusive categories, we use the background reported in  Ref.~\cite{Chatrchyan:2012tw} for the $jj$ class and 
Ref.~\cite{CMS-PAS-HIG-12-002} for the leptonic one.
The final number of  background events is obtained by performing a simple cut on $m(\gamma\gamma)$ around
the Higgs mass (120 GeV), consistent with the expected
CMS mass resolution, which corresponds to a $\pm 3\,$GeV window
for barrel-barrel photon categories and for the exclusive
$jj$ and leptonic categories, and a $\pm 6\,$GeV window for
photon categories with at least a photon in the endcap. The background is thus obtained by integrating 
the number of events estimated from data in these windows.

Since we want to scan the $(a,c)$ plane, signal efficiencies for each
category and for each of the different Higgs production mechanisms are needed. We 
use Montecarlo generators to determine these efficiencies. For gluon-gluon fusion and VBF we use POWHEG at 
next-to-leading order (NLO) \cite{Alioli:2008tz,Nason:2009ai},
while for VH we use PYTHIA at leading order (LO). 
The sum of the contributions from the different production
mechanisms are then scaled to give the total number of Higgs events
in the 4 photon categories and in the $jj$ category as reported in
\cite{Chatrchyan:2012tw} and in the leptonic category as reported in \cite{CMS-PAS-HIG-12-002}.
We assume that the efficiency of the $m(\gamma\gamma)$  cut described above 
is approximately 100\% on the signal.

We derive our results for three different analyses: 
\begin{itemize}
\item[--] one with 4 categories based on $R_9$ and photon pseudorapidity variables, which makes no use of the $p_T(\gamma\gamma)$ spectrum,
as in~\cite{CMS-PAS-HIG-11-030}; 
\item[--] one with 8 categories based on $R_9$, photon pseudorapidity and $p_T(\gamma\gamma)$ variables to help discriminating between different production mechanisms, thanks to the harder $p_T$(Higgs) in VBF and VH mechanisms compared to gluon-gluon fusion; 
\item[--] one with 8 inclusive plus two exclusive ($jj$ and $1l$) categories, to fully exploit the physics potential.
\end{itemize}
A summary of the number of background and SM signal events expected per
fb$^{-1}$ is reported in Tab.~\ref{tab:xsecsEXCL} and Tab.~\ref{tab:xsecsINCL} for the last two analyses.
In the case of the 4-category analysis, the number of events in  each of the $R_9$ and $\eta$ classes
is obtained from Tab.~\ref{tab:xsecsINCL}  by summing together the  corresponding  high and low  $p_T(\gamma\gamma)$ events.
%
\begin{table}[t]
\begin{center}
{\small
\begin{tabular}{lcccccccccc}
 &  & &  \multicolumn{4}{c}{$p_T(\gamma\gamma) < 40\,$GeV} &  \multicolumn{4}{c}{$p_T(\gamma\gamma) > 40\,$GeV} \\[0.1cm]
 &  &       & $R_{9}^>$ &  $R_{9}^<$ &  $R_{9}^>$ & $R_{9}^<$ & $R_{9}^>$ & $R_{9}^<$ & $R_{9}^>$ &  $R_{9}^<$ \\[-0.15cm]
 & $1l$ & $jj$ & BAR &  BAR &  END & END &  BAR & BAR & END &  END \\
\hline \\[-0.45cm]
 GGF & 0 & 0.14 & 3.23 & 3.40 & 1.20 & 1.44 &  1.55 & 1.64 & 0.58 & 0.69 \\ 
 VBF  & 0 & 0.44 & 0.067 & 0.071 & 0.026 & 0.031   & 0.17 & 0.18 & 0.066 & 0.079  \\
 VH & 0.089 & 0.0035 & 0.059 & 0.063 & 0.028 & 0.033  & 0.17 & 0.18 & 0.081 & 0.097 \\[0.2cm]
 background & 0.25 & 2.88 & 85.4 & 126 & 134 & 188  & 36.4 & 53.7 & 57.7 & 80.3 
\end{tabular}
}
\caption[]{\label{tab:xsecsEXCL}
\small
Number of events (per fb$^{-1}$) in each of the 10 categories of the exclusive analysis for the 
signal in the SM (for each Higgs production mode) and total background.
}
\end{center}
\end{table}
%
\begin{table}[t]
\begin{center}
{\small
\begin{tabular}{lcccccccc}
  &  \multicolumn{4}{c}{$p_T(\gamma\gamma) < 40\,$GeV} &  \multicolumn{4}{c}{$p_T(\gamma\gamma) > 40\,$GeV} \\[0.1cm]
        & $R_{9}^>$ &  $R_{9}^<$ &  $R_{9}^>$ & $R_{9}^<$ & $R_{9}^>$ & $R_{9}^<$ & $R_{9}^>$ &  $R_{9}^<$ \\[-0.15cm]
  & BAR &  BAR &  END & END &  BAR & BAR & END &  END \\
\hline \\[-0.45cm]
 GGF  & 3.21 & 3.41 & 1.19 & 1.43 & 1.61 & 1.71 & 0.60 & 0.72   \\ 
 VBF  &  0.091 & 0.096 & 0.031 & 0.036 & 0.31 & 0.33 & 0.10 & 0.13 \\ 
 VH & 0.067 & 0.070 & 0.030 & 0.036 & 0.20 & 0.21 & 0.089 & 0.11 \\[0.2cm]
 background & 85.8 & 126 & 135 & 189 & 36.6 & 53.9 & 58.0 & 80.6 
\end{tabular}
}
\caption[]{\label{tab:xsecsINCL}
\small
Number of events (per fb$^{-1}$) in each of the 8 categories of the inclusive analysis for the 
signal in the SM (for each Higgs production mode) and total background.
}
\end{center}
\end{table}
%
Starting from the number of signal events predicted in the SM for each production mode, the number of events 
for arbitrary couplings $a$, $c$ is easily obtained by rescaling the Higgs production cross sections
and partial decay rates, as detailed in the Appendix.
For each category $i$, given the number of signal ($n^i_s(a,c)$), background ($n^i_b$) and observed events ($n^i_{obs}$), we construct a 2D posterior probability 
\begin{equation}
p(a,c | n^i_{obs}) = p(n^i_{obs} | n^i_s(a,c)+n^i_b) \times \pi(a,c)
\end{equation}
following the  Bayesian approach.~\footnote{See for example Ref.~\cite{D'Agostini:2003nk} for a primer.}
The total probability is then obtained as the product of the single probabilities.
The likelihood function $p(n^i_{obs} | n^i_s+n^i_b) $ is modeled by a Poisson distribution, and we take a flat prior $\pi(a,c)$ on the square $-3 \leq a,c \leq +3$ 
(vanishing outside) as done in Ref.~\cite{Azatov:2012bz}. 
The effect of systematic uncertainties on the signal is taken into account by letting the fraction of signal events in each category and 
from each production mode fluctuate.
We do so by varying all the fractions with a single nuisance parameter~$\theta_s$, 
so that $n^i_s \to n^i_s (1+\theta_s)$, except for the GGF fraction in the 
$jj$ category which is varied with a different parameter $\theta_s^{GGFjj}$. 
The total probability is then marginalized over $\theta_s$ and $\theta^{GGFjj}_s$, which are taken to be distributed with a truncated Gaussian with zero mean and
standard deviation equal to respectively $\Delta \theta_s = 0.15$ and $\Delta \theta^{GGFjj}_s = 0.70$.
This corresponds to treating the systematic errors on the signal as 100\% correlated in all categories and production modes, which is a reasonable approximation 
considering that the largest uncertainty comes from the theoretical prediction of the Higgs production cross sections, except for the GGF events in the $jj$ category,
whose largest uncertainty originates from the efficiency of the kinematic cuts applied~\cite{Chatrchyan:2012tw}. 
We neglect all systematic uncertainties on the background.

To check that all assumptions, efficiency estimates and statistical analysis
are reasonable and robust, we derived the  expected limits on the signal strength modifier  considered by CMS and ATLAS, $\mu=\sigma\times BR/(\sigma \times BR)_{SM}$,
in two different scenarios: in the SM hypothesis, where the limit is extracted with 4 categories to mimic the CMS analysis of \cite{CMS-PAS-HIG-11-030}, 
and in the Fermiophobic (FP) scenario, where the limit is extracted with  8+2 categories to mimic the CMS analysis of \cite{CMS-PAS-HIG-12-002}. 
To do so we set $a = c = \sqrt{\mu}$ and $n_{obs} = n_b$ in our likelihoods, and use a flat prior on $\mu$ for $\mu >0$ (zero otherwise) as adopted by ATLAS 
and CMS~\cite{LHCsummer}.
For a $120\,$GeV Higgs we find the following 95\% expected limits: $\mu^{95\%} = 1.4$ for the SM case,  to be compared with $1.6$ reported by CMS;
$\mu^{95\%} = 0.36$ for the FP case,  to be compared with $0.30$ reported by CMS.
Both estimates are in reasonable agreement with the official value, considering the approximations done in our method.
The looser limit in the FP case is probably due to the use of two  $p_T(\gamma\gamma)$ categories instead of the full 2D fit approach
(m($\gamma\gamma$) vs $p_T(\gamma\gamma)$)  performed in \cite{CMS-PAS-HIG-12-002}.

\section{Results}

We now present the results of our analysis of the $\gamma\gamma$ channel for a Higgs mass of $120\,$GeV.
As discussed in more details in the following, at the qualitative level our results apply reasonably well to the range of Higgs masses
$m_h = 120-130\,$GeV, while at the quantitative level differences can become important for $m_h \gtrsim 125\,$GeV.

\begin{figure}[tb]
\begin{center}
\includegraphics[width=0.58\textwidth,clip,angle=0]{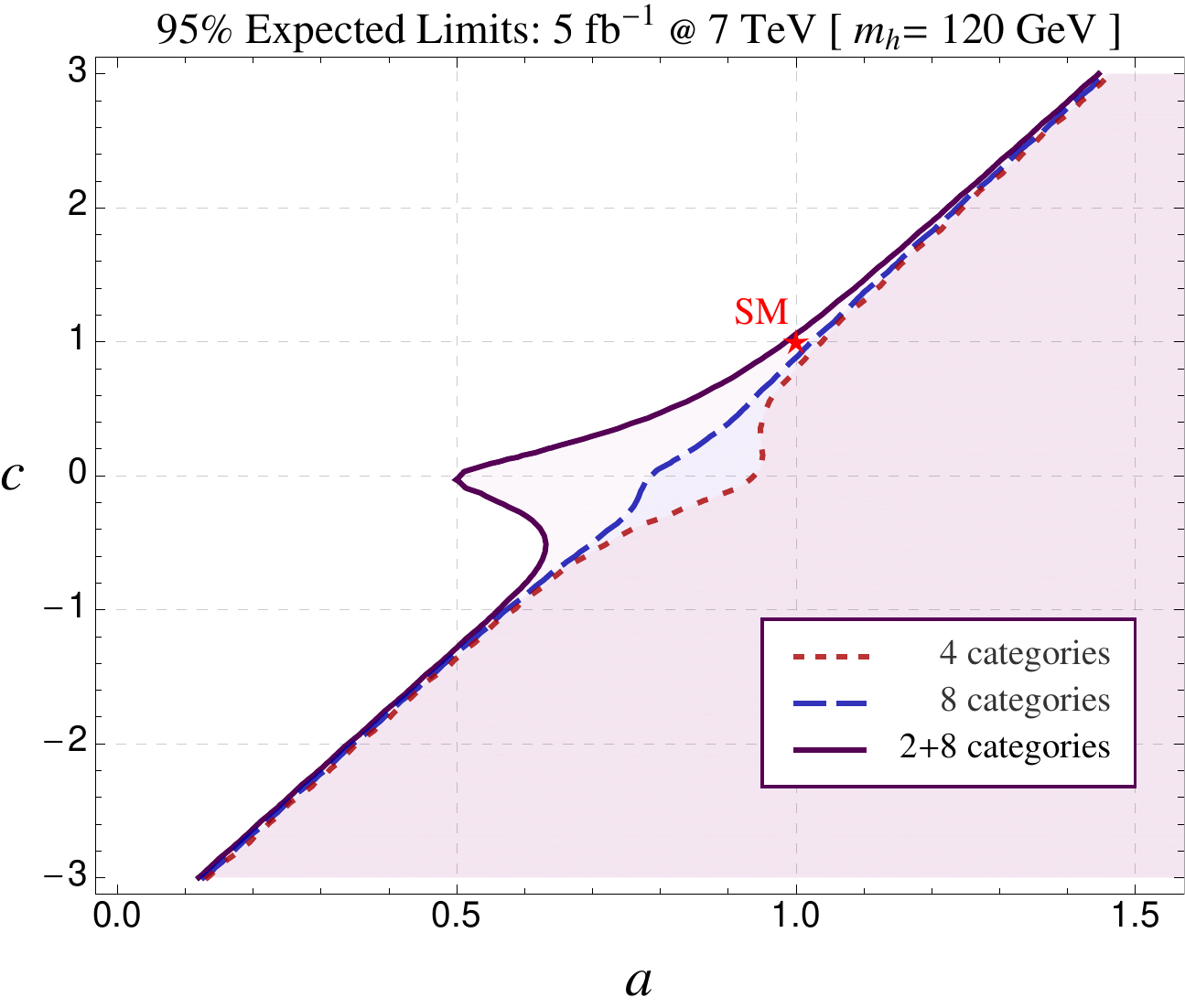}
\caption[]{\small
Expected exclusion limits from $\gamma\gamma$ at $\sqrt{s} = 7\,$TeV with $L=5\,\text{fb}^{-1}$ and $m_h=120\,$GeV.
Purple solid curve: exclusive analysis with 8+2 categories; Dashed blue curve: inclusive analysis with 8 categories; 
Dotted red curve:  inclusive analysis with 4 categories. The area on the right of each curve is excluded at $95\%$ probability.
\label{fig:95limits}
}
\end{center}
\end{figure}

We start with a discussion of the expected $95\%$ exclusion limits in the $(a,c)$ plane, which are shown in Fig.~\ref{fig:95limits} for a center-of-mass energy 
$\sqrt{s} = 7\,$TeV and an integrated luminosity $L =5\,\text{fb}^{-1}$, approximately the amount of luminosity accumulated individually by ATLAS and CMS in 2011.
%
%
One can see that the fully-exclusive analysis with 10 categories (purple solid curve) is much more powerful in the
$c\sim 0$ region compared to the  inclusive analysis with 4 categories (dotted red curve), \textit{e.g.} the one performed by CMS in 
Ref.~\cite{CMS-PAS-HIG-11-030}.
For $c\to 0$ the Higgs couplings to fermions vanish and the total production cross section, which for large 
values of $|c|$ is strongly dominated by  gluon fusion, receives its main contribution from VBF and $W/Z$ associated production.
An enhanced sensitivity to these production modes, as obtained by including the two exclusive event classes, can thus lead to much stronger constraints.
An appreciable, though milder improvement on the limit is also obtained in the vicinity of the SM point, in agreement with the results 
of Ref.~\cite{Chatrchyan:2012tw}.~\footnote{We note in passing that for $a=c$ the constraint from the 4-category analysis is stronger than the limit on the 
signal modifier $\mu^{95\%} = 1.6$ discussed in the previous section. Indeed, it can be shown that a 2D probability with flat prior gives the same
limit on the line $a=c$ of a 1D probability with non-flat prior on $\mu$.}

Interestingly, a further subdivision of the 4 inclusive categories into two sets with respectively large and small 
$p_T(\gamma\gamma)$ also increases the sensitivity in the fermiophobic region (dashed blue curve). 
This is because the distribution of the transverse
momentum of the $\gamma\gamma$ pair tends to be harder for events produced through VBF and associated production,  so that requiring larger
values of  $p_T(\gamma\gamma)$ increases the relative importance of these production modes compared to gluon fusion. An analysis with 8 categories was performed 
by CMS in 2011 on 1.66 fb$^{-1}$ of data (a subset of the total 2011 data set) and is reported in Ref.~\cite{CMS-PAS-HIG-11-021}.
We find, although the corresponding curve is not shown in Fig.~\ref{fig:95limits}, that once the two exclusive categories optimized respectively for VBF 
and associated production are included in the analysis, having 8 additional `inclusive' categories instead of 4 does not appreciably improve the sensitivity in the
$(a,c)$ plane. In other words, performing an exclusive analysis with 4+2 categories leads to constraints on the couplings $a$, $c$ quite similar to those
obtained with our analysis which makes use of 8+2 categories. This in fact agrees with the naive expectation, considering that the fraction of events produced 
through VBF and associated production that fall into the inclusive categories is quite small: see Table~\ref{tab:xsecsEXCL}.
To summarize, we find that an exclusive analysis of $h\to\gamma\gamma$
is more powerful than an inclusive one to set limits on the Higgs couplings, especially in regions where the importance of the VBF and associated
production modes is enhanced compared to gluon fusion.

A fully exclusive analysis of the $\gamma\gamma$ channel is even more useful once the observation of a signal has been established and 
it comes to extracting the Higgs couplings. We illustrate this in the following
by showing contours of equal probability in the plane $(a,c)$ obtained by injecting a specific signal and assuming $L= 20\,\text{fb}^{-1}$
with $\sqrt{s} = 7\,$TeV. This should be a reasonable approximation of the data set which will be individually accumulated in 2012 by ATLAS and CMS 
at $\sqrt{s} = 8\,$TeV. We have chosen to perform our simulations at $7\,$TeV (rather than 8) to 
facilitate comparison with the previous results and to be conservative since
it is still not clear what the real performances of the detectors will be with the larger pile-up rate.

\begin{figure}[tb]
\begin{center}
\includegraphics[width=0.48\textwidth,clip,angle=0]{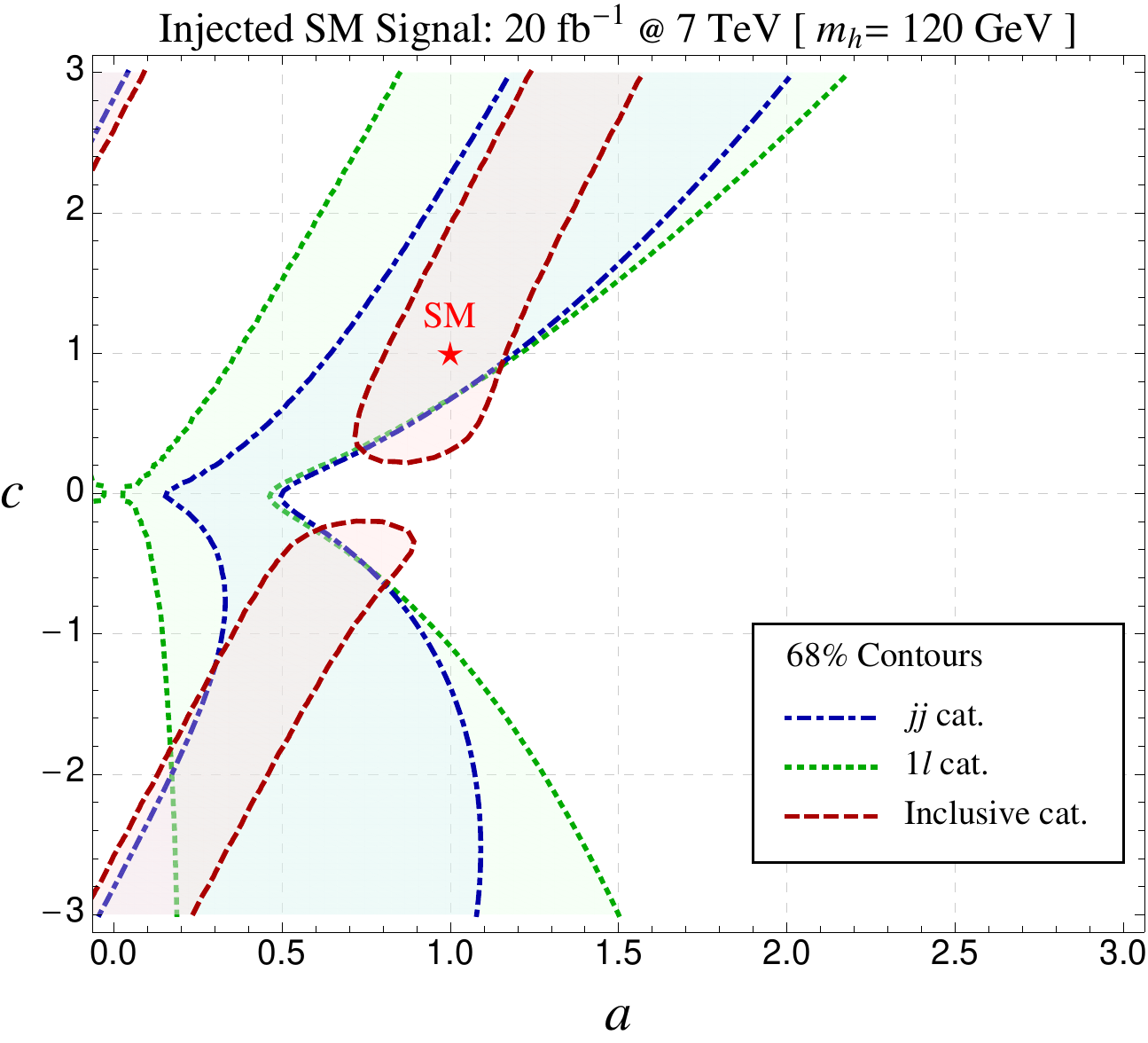}
\includegraphics[width=0.48\textwidth,clip,angle=0]{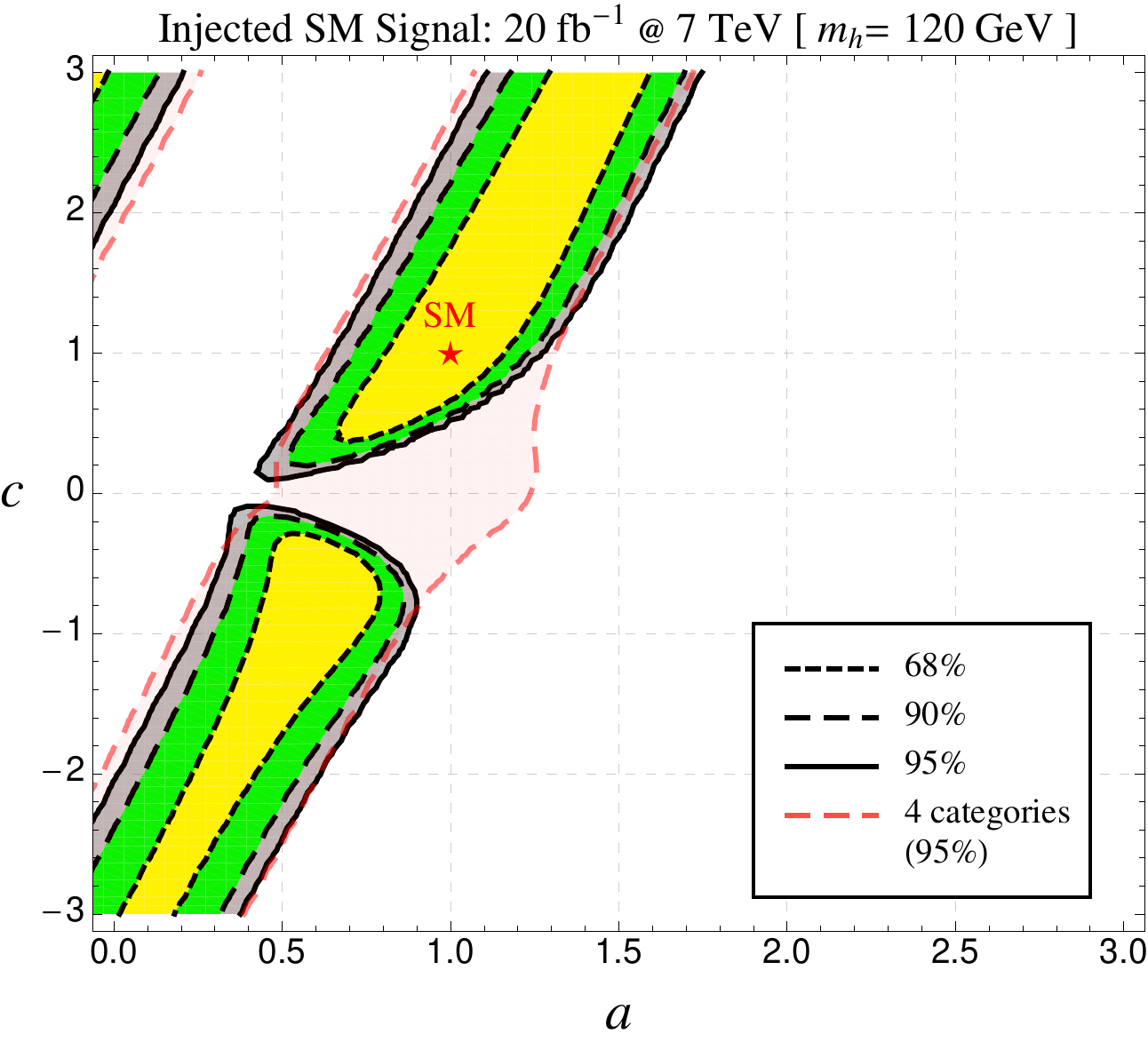}
\caption[]{\small
Contours of constant  probability for $\gamma\gamma$ in the plane $(a,c)$ obtained by injecting the SM signal $(a=1,c=1)$.
Left plot: 68\% contours for the $jj$, $1l$ and inclusive categories.
Right plot:  68\%, 90\%, 95\% contours in the exclusive analysis with 8+2 categories and 95\% contour in the inclusive analysis
with 4 categories.
Both plots are for $\sqrt{s} = 7\,$TeV with $L=20\,\text{fb}^{-1}$ and $m_h=120\,$GeV.
\label{fig:SMinjected}
}
\end{center}
\end{figure}

Figure~\ref{fig:SMinjected} illustrates the case of an injected SM signal ($a=1, c=1$). 
%
%
The plot on the left shows the $68\%$ probability contours selected by respectively the $jj$, $1l$ and (the combination of the eight) inclusive categories.
Related results were discussed in Refs.~\cite{Lafaye:2009vr,Carmi:2012yp,Azatov:2012bz,Espinosa:2012ir,Giardino:2012ww,Ellis:2012rx}, although following 
different approaches and assumptions than ours.
The shape of the various regions can be easily reproduced considering that the yield of the
two exclusive categories is dominated respectively by events produced via VBF and associated production, while the inclusive
categories are dominated by gluon fusion. Defining the ratio
\begin{equation}
\mu_i = \frac{\sigma_i \times BR(\gamma\gamma)}{[\sigma_i \times BR(\gamma\gamma)]_{SM}}
\end{equation}
as the yield in a given category $i$ in SM units, it thus follows
\begin{equation}
\label{eq:ratios}
\mu_{jj} \sim \mu_{1l} \sim a^2 \, \frac{(4.5\, a - c)^2}{c^2} \, , \qquad
\mu_{incl} \sim ( c^2 + \zeta\, a^2) \, \frac{(4.5\, a - c)^2}{c^2} \, ,
\end{equation}
where the factor $(4.5\, a - c)^2$ follows from the branching ratio to $\gamma\gamma$, and $\zeta$ parametrizes the small 
contamination of  VBF and VH events in the inclusive categories.
Eq.~(\ref{eq:ratios}) reproduces to good accuracy the shape of the different regions of Fig.~\ref{fig:SMinjected}.
In particular, the non-negligible contribution of VBF and VH events in the inclusive categories with high $p_T(\gamma\gamma)$ (see Table~\ref{tab:xsecsEXCL}) 
removes the long tail at large $a$  and small $c$ of the area which would be selected by the remaining four inclusive classes with low  $p_T(\gamma\gamma)$.
\footnote{See for example the upper right plot in Fig.~2 of Ref.~\cite{Carmi:2012yp}, where the contribution of VBF and VH events to the inclusive categories was neglected.}
The resulting 68\% region selected by the combination of all inclusive categories is that shown in red in the left plot of Fig.~\ref{fig:SMinjected}, which 
stretches along the line $(4.5\, a - c) = const.$ passing through the SM point. We have checked, on the other hand, that the contamination of GGF events in the $jj$
category modifies only marginally the shape of the 68\% probability region selected by this category.

For $c\to 0$ the exclusive $jj$ and $1l$ categories favor values $a < 1$,  which ensure a suppression of the production
cross section and compensate the strong increase in the branching ratio, as required to reproduce $\mu_{jj,1l} \sim 1$.
On the contrary, the region $c\sim 0$ is disfavored for any value of $a$ by the inclusive categories, since their yield is strongly suppressed in the fermiophobic
limit. As a result, by injecting the SM signal, an exclusive analysis of $h\to\gamma\gamma$ can exclude the fermiophobic region $c\simeq 0$ with a probability
of more than 95\%; see the plot on the right in Fig.~\ref{fig:SMinjected}. This is especially true for the benchmark point $(a=1, c=0)$, which predicts
too many events in the $jj$ and $1l$ categories and too few in the inclusive ones.
On the other hand, it is not possible to exclude this point and the region surrounding it 
by means of a 4-category inclusive analysis; see the dashed red curve in the same plot.
Indeed, the total $\gamma\gamma$ yield for $(a,c) \sim (1,0)$ is approximately that of the SM (see for example the discussion in Ref.~\cite{Gabrielli:2012yz}), and 
the overall sensitivity decreases as a consequence of the absence of the clean exclusive categories.

In order to derive an estimate of how the results in Fig.~\ref{fig:SMinjected} change with the Higgs mass, we have repeated our analysis by varying $m_h$ and
assuming that the background yield and the selection efficiencies do not change significantly. This is expected to be a reasonably accurate approximation for  
$m_h = 120-130\,$GeV. In this range of masses the variation of the signal yield is driven by the change in the Higgs production cross sections and in the 
$\gamma\gamma$ branching ratio,  with the latter giving the dominant effect. We find that even for $m_h = 130\,$GeV the contours of Fig.~\ref{fig:SMinjected} 
are only slightly modified. This is because for $(a=1, c=1)$ the signal yield, hence the injected one, changes by less than $\sim 15\%$.
The larger distortion occurs  in the fermiophobic region $c\sim 0$, where the $\gamma\gamma$ branching ratio is enhanced, which is however largely disfavored
by combining the inclusive and exclusive categories. 
We thus conclude that our results hold with good accuracy in the range  $m_h = 120-130\,$GeV.

The exclusive analysis selects two regions with high probability: one includes the SM point, the other corresponds to negative values
of $c$ (yellow areas in the right plot of Fig.~\ref{fig:SMinjected}). The presence of a second solution in addition to $(a,c) = (1,1)$ is a direct consequence
of the quadratic dependence of the yields in eq.(\ref{eq:ratios}) on $a$, $c$ and the interference of the 1-loop top and $W$  contributions to the $\gamma\gamma$ decay rate: 
by injecting a given signal $(a_0, c_0)$, there is a second solution
\begin{equation}
\label{eq:secondsol}
a \simeq a_0 \, \frac{4.5\, a_0 - c_0}{4.5\, a_0 + c_0}\, , \qquad 
c \simeq - c_0 \, \frac{4.5\, a_0 - c_0}{4.5\, a_0 + c_0}\, , 
\end{equation}
which gives the same yields $\mu_{jj}$, $\mu_{1l}$ and $\mu_{incl}$. For $(a_0 ,c_0) = (1,1)$ the second solution corresponds to $(0.64, -0.64)$, which is
indeed the position of the second maximum of the 2D probability whose contours are shown in Fig.~\ref{fig:SMinjected}.

\begin{figure}[tb]
\begin{center}
\includegraphics[width=0.48\textwidth,clip,angle=0]{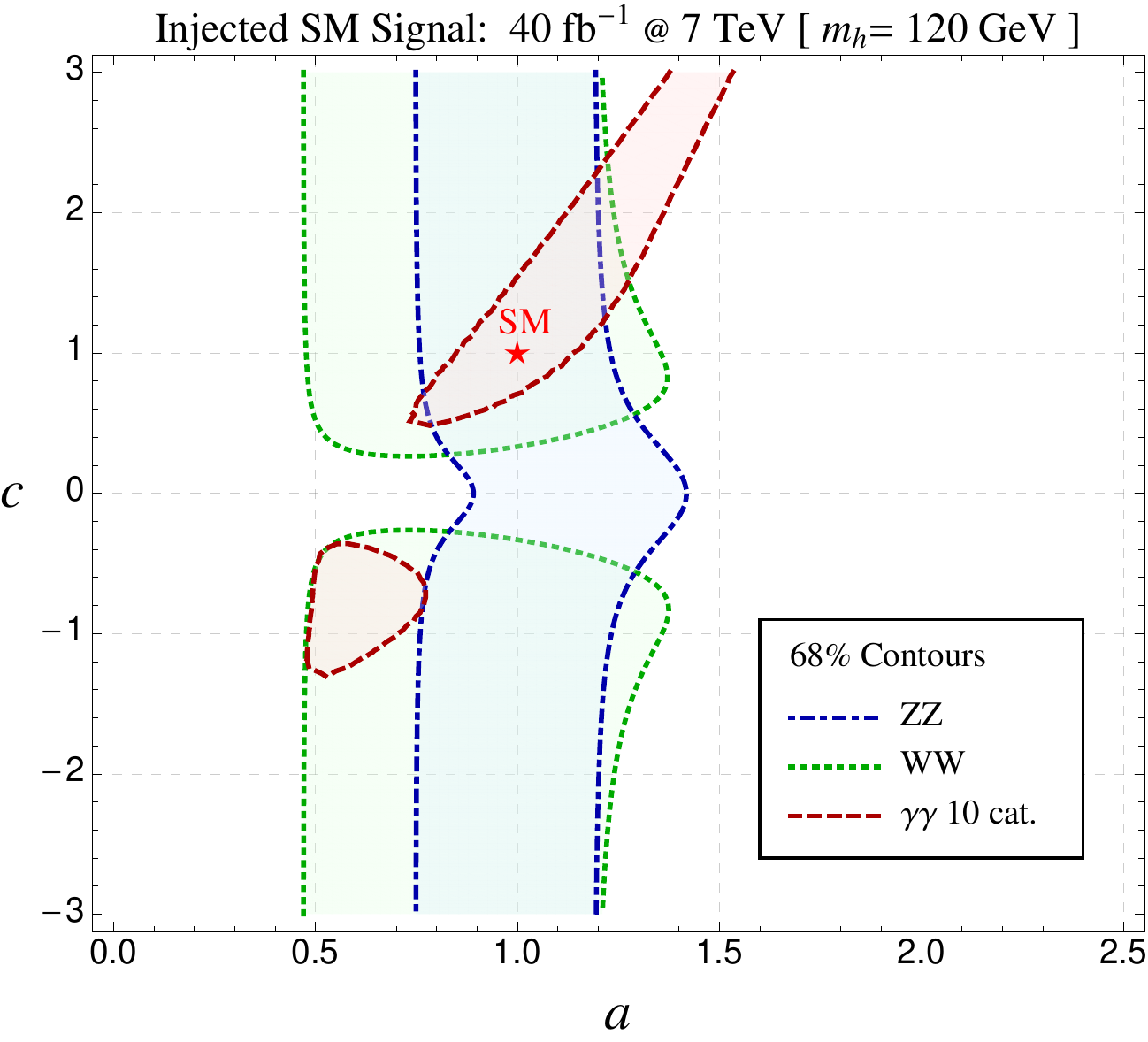}
\includegraphics[width=0.48\textwidth,clip,angle=0]{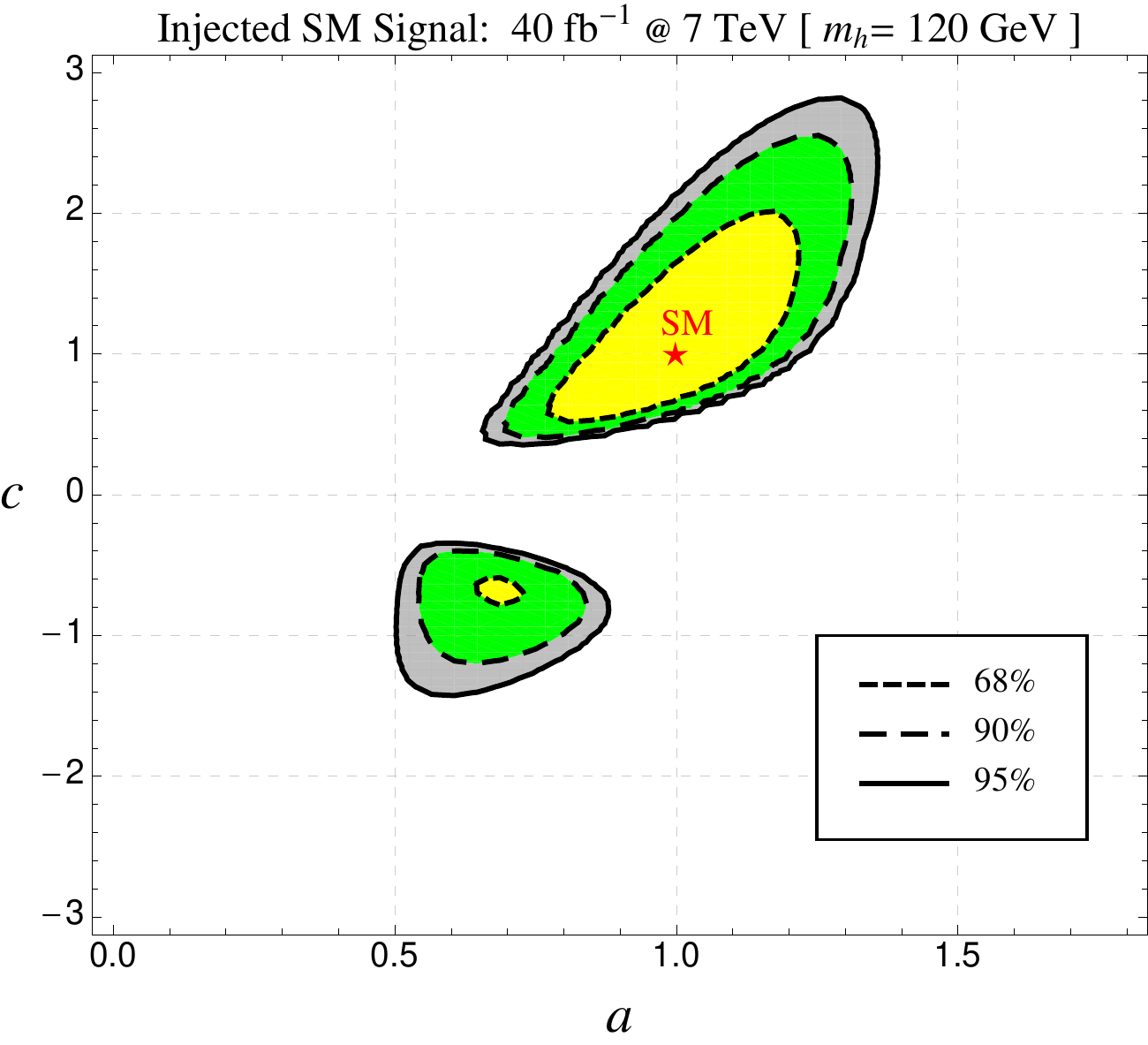}
\caption[]{\small
Contours of constant probability in the plane $(a,c)$ for $\gamma\gamma$, $ZZ$ and $WW$ obtained by injecting the SM signal $(a=1,c=1)$.
Left plot: 68\% contours for individual $\gamma\gamma$ (10-categories exclusive analysis, red area), $WW\to l\nu l\nu$ (5-categories exclusive analysis, green area)
and $ZZ\to4l$ (inclusive analysis, blue area) channels. Right plot: 68\%, 90\%, 95\% contours for their combination. For $WW$ and $ZZ$ the probability function has
been constructed by rescaling the number of events reported by CMS respectively in Ref.~\cite{Chatrchyan:2012ty} and Ref.~\cite{Chatrchyan:2012dg}; see text.
\label{fig:WWZZgaga}
}
\end{center}
\end{figure}

The existence of a second degenerate solution in the plane $(a,c)$ was noticed and discussed in 
Refs.~\cite{Lafaye:2009vr,Azatov:2012bz,Espinosa:2012ir,Giardino:2012ww,Ellis:2012rx}. 
Breaking such degeneracy will require large integrated luminosity and the combined use of several channels. 
An extrapolation of the results of the current searches to higher luminosity indicates that the most sensitive channels in this regard 
are $\gamma\gamma$ and $ZZ\to 4l$, while others, like $WW$ and $\tau\tau$, are less powerful.
Although performing  an exclusive analysis for each decay channel will play a crucial role also in this case, a complete resolution
of the degeneracy might require considering more refined strategies. This is for example illustrated by Fig.~\ref{fig:WWZZgaga}, where we show the probability contours
obtained at $L= 40\,\text{fb}^{-1}$ (the total amount of integrated luminosity which might be obtained by the end of 2012 by CMS and ATLAS together) 
from $\gamma\gamma$, $ZZ\to 4l$ and $WW\to l\nu l\nu$ (left plot) and their combination (right plot). 
%
%
For the $WW$ channel we have considered the exclusive analysis performed
by CMS~\cite{Chatrchyan:2012ty} for $m_h = 120\,$GeV 
(see Ref.~\cite{Azatov:2012bz} for details). In the case of $ZZ$ we have performed a simple cut-and-count analysis by considering the number of signal 
and background events expected by CMS in a $\pm 5\,$GeV window around $m(4l) = 120\,$GeV.~\footnote{More specifically we used the 
right plot of Fig.~2 of Ref.~\cite{Chatrchyan:2012dg} and summed the number of  
events in five bins around $m(4l) = 120\,$GeV. In this way we find $n_s^{SM} =1.5$, $n_b =1.7$ 
respectively for the number of SM signal and background events with $L=4.7\,\text{fb}^{-1}$.
Since the CMS analysis is inclusive, we have rescaled the SM yield in the plane $(a,c)$ by assuming that the cut efficiencies
are the same for each of the various production modes. Although this is known to be a very rough approximation, it is the best one can do in absence
of more detailed information.}
We have constructed the posterior probability by including a $15\%$ systematic error on the signal, while we have neglected the systematic uncertainty on the
background since this is expected to be small for a shape-based analysis like $ZZ\to 4l$ (and similarly $\gamma\gamma$) once sufficient statistics has been accumulated.

As the left plot of Fig.~\ref{fig:WWZZgaga} illustrates, the projected sensitivity of the current $WW$ analysis to $L= 40\,\text{fb}^{-1}$ is poor and does not
help much to remove the second solution. This is due in large part to the effect of the systematic uncertainties, which  are large for $WW$.
 It is not clear if this systematic error will be reduced in a future analysis or if it will increase as due to the larger
uncertainty on the missing energy measurement which could follow from the higher pile-up rate at 8 TeV.  A slightly enhanced sensitivity for
$WW$ is expected if the Higgs is heavier than $120\,$GeV, as the result of the increase in the corresponding branching ratio.
The $ZZ\to 4l$ channel, on the other hand, is much more clean and has a strong impact in disfavoring the second solution. After its inclusion
in the fit, the peak of the probability at $a = - c = 0.64$ is $\sim 5$ times smaller than the peak at $a=c=1$ (see the right plot of Fig.~\ref{fig:WWZZgaga}).
We have checked that the $\tau\tau$ channel selects a broad region in the $(a,c)$ plane, and it has very little impact on the global fit.~\footnote{We have 
made a very crude cut-and-count estimate based on the CMS analysis of Ref.~\cite{Chatrchyan:2012vp}. We find that assuming a 10\% systematic uncertainty 
on the background the precision on the $(a,c)$ plane is very poor. A more refined result would require a detailed analysis which is beyond the
scope of this paper.} For this reason we have not included it in Fig.~\ref{fig:WWZZgaga}. 
In this regard our results do not agree with the early analysis of Ref.~\cite{duhrssen}, which used a much more optimistic estimate of the background and 
found that $\tau\tau$ was one of the most sensitive channels for $m_h = 120\,$GeV.~\footnote{A similar underestimation of the background for $\tau\tau$ 
is present also in the analyses of Refs.~\cite{Duhrssen:2004cv,Lafaye:2009vr}. See also the discussion and results of  Ref.~\cite{Rauch:2012wa} on 
the expected sensitivity on the Higgs couplings obtained by making use of the actual background estimates and errors reported in the current experimental 
analyses as compared to earlier Montecarlo studies.}

Our results show that by extrapolating the current analyses to 40 fb$^{-1}$ the second solution can be disfavored but not completely eliminated.
A complete removal of the degeneracy will require more integrated luminosity or substantial improvements of the present analyses, possibly following from
new strategies. The use of ratios of yields in  different categories within the same decay channel or different channels, as recently suggested by Ref.~\cite{Espinosa:2012ir}
as a way to reduce the degeneracy, does not seem to provide a resolution in this case. Its main advantage indeed is that it helps to reduce the systematic uncertainties,
which are however already expected to be  small for $\gamma\gamma$ and $ZZ\to 4l$.
We find that by setting to zero the systematic error on the signal of both $\gamma\gamma$ and $ZZ$ the contours of Fig.~\ref{fig:WWZZgaga} are marginally
modified. In particular, the second solution becomes excluded at $68\%$ but the extension of the $90\%$ and $95\%$ probability regions is only slightly reduced.
Concentrating on the  solution centered at the SM point, the plot of Fig.~\ref{fig:WWZZgaga} suggests that with 40 fb$^{-1}$, if the Higgs is that of the SM,  the coupling $a$
can be measured with a precision of $\sim 25\%$, while the uncertainty on $c$ is of the order of $100\%$.
Our estimate for $a$ seems to be in agreement with the recent results of \cite{Rauch:2012wa}, which however reports a significantly 
smaller uncertainty on $c$.

\begin{figure}[tbp]
\begin{center}
\includegraphics[width=0.47\textwidth,clip,angle=0]{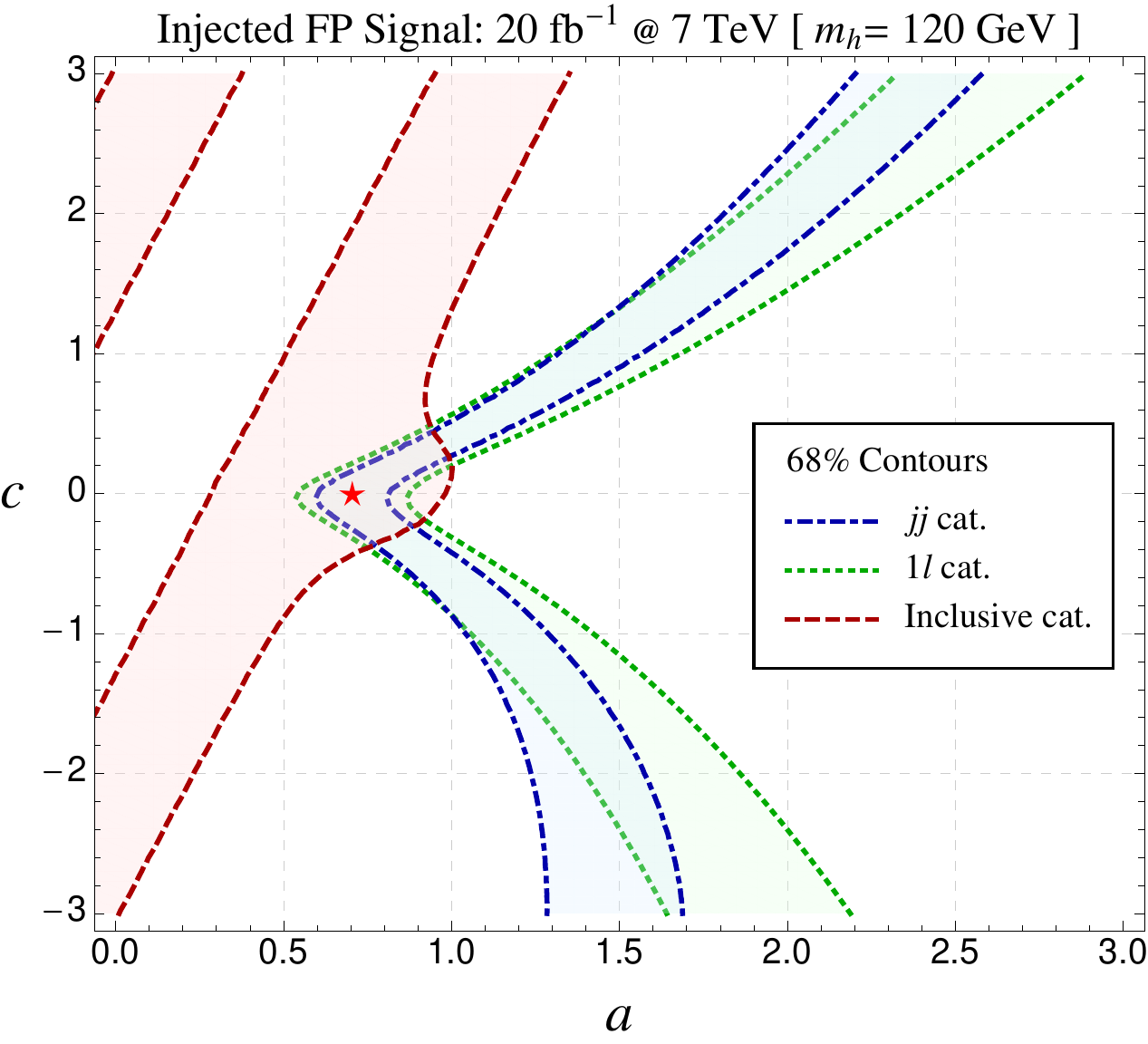}
\includegraphics[width=0.48\textwidth,clip,angle=0]{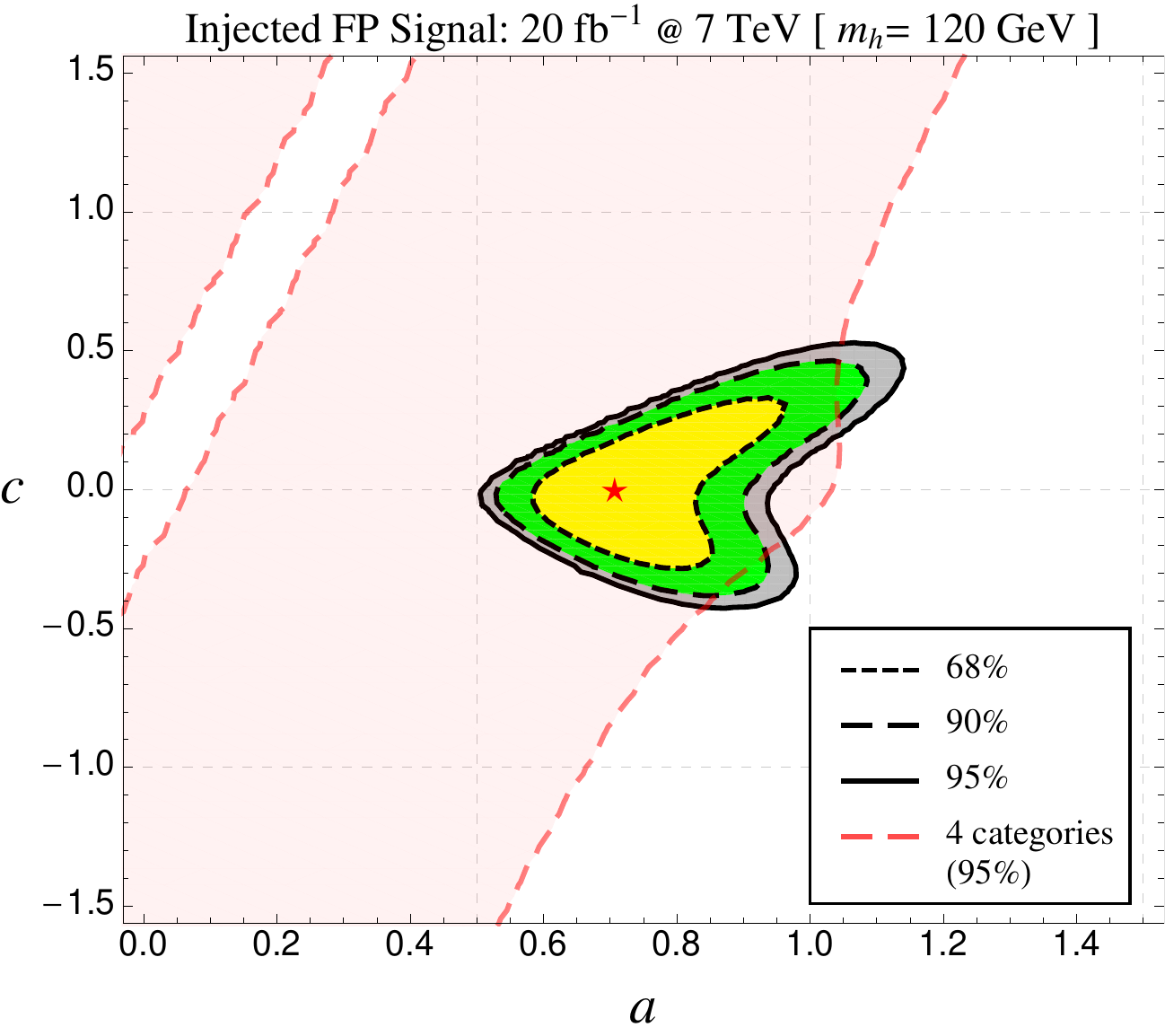}
\caption[]{\small
As for Fig.~\ref{fig:SMinjected} with injected fermiophobic signal $(a=1/\sqrt{2},c=0)$.
\label{fig:FPinjected}
}
\end{center}
\end{figure}

We end this section by showing in Fig.~\ref{fig:FPinjected} the contours of equal probability for an injected signal $(a=1/\sqrt{2}, c=0)$, for $L= 20\,\text{fb}^{-1}$.
We choose this point as representative of a fermiophobic scenario since it is realized in the composite Higgs model MCHM5~\cite{Contino:2006qr}
and it was already considered in previous works. Notice that although for $m_h = 120\,$GeV  such choice of couplings is excluded at 95\% CL 
by the current CMS combined  results~\cite{CMS-PAS-HIG-12-008}, it is still allowed for $123\,\text{GeV} < m_h < 130\,\text{GeV}$.~\footnote{By 
comparison, the `standard' benchmark point $(a=1,c=0)$ is excluded at 95\% CL in the whole range $110-192\,$GeV~\cite{CMS-PAS-HIG-12-008}.}
%
%
As expected from Eq.~(\ref{eq:secondsol}), in this case there is no degeneracy of solutions. By performing an exclusive analysis,
the maximum of the probability is obtained in a small region  around $(1/\sqrt{2},0)$ where  $\mu_{incl}$ is small and $\mu_{jj,1l} \sim 5$.
The Higgs couplings  $a$, $c$  can be  determined  in this case  with 
a precision of $\sim 35\%$. On the other hand, an inclusive analysis with 4 categories is dramatically less powerful and selects only a broad region in the plane
(red area in the right plot of Fig.~\ref{fig:FPinjected}).
We checked that the same qualitative conclusions apply for $m_h = 125\,$GeV, although the uncertainty on the couplings increases to $\sim 45\%$.
On the other hand, for larger Higgs masses the contours of Fig.~\ref{fig:FPinjected}  become quickly broader, and already at $m_h =130\,$GeV the 90\% region of
the combined fit forms an open strip in the plane. This is mostly due to the decrease of the injected yield implied by the fast drop of the $\gamma\gamma$
branching ratio at  heavier Higgs masses for $c=0$. We thus conclude that while our results for the fermiophobic case apply reasonably well up to
$m_h = 125\,$GeV, assessing the precision on the Higgs couplings at larger Higgs masses will require a dedicated analysis.

\section{Conclusions}
\label{sec:Conclusions} 

If a Higgs-like scalar is discovered at the LHC in 2012, we will enter a new exciting phase in which the main
focus will be on determining its couplings as precisely as possible. This will eventually shed light on the dynamics
behind EWSB, confirming or excluding the SM paradigm.
It is clear that a correct theoretical interpretation of the experimental results will be crucial in achieving  this goal, but much will also
depend on the way the  analyses are performed and the results presented by the experimental collaborations.
This is evident even now, considering the numerous theoretical papers appearing
recently~\cite{Carmi:2012yp,Azatov:2012bz,Espinosa:2012ir,Giardino:2012ww,Rauch:2012wa,Ellis:2012rx} which try to interpret the 
results from the LHC searches on the SM Higgs in terms of models of new physics.
In fact, a fully correct interpretation is not always possible because the published experimental papers often do not contain
sufficient information.
For example, even a crude estimate of the impact of the LHC searches on generic models of new physics
requires knowing the cut efficiencies for each of the production modes and each of the event categories. Hopefully, this information will be made public in the future
by the experimental collaborations. However, even knowing that, a rigorous determination of the Higgs couplings is possible only by including all the
systematic effects and correlations among different categories and different channels. Although a few theoretical groups have taken the challenge seriously
(for example the SFitter collaboration~\cite{Lafaye:2004cn,Rauch:2012wa}), the experimental collaborations themselves seem to be those who can more easily confront 
this formidable task.
The main obstacle in this regard is that performing an analysis necessarily requires \textit{some} assumption about which (class of) models are under scrutiny.
This highlights the importance of a consistent theoretical framework that allows one to explore the largest possible landscape of theories with the smallest number
of assumptions.

The electroweak chiral Lagrangian introduced in~\cite{Contino:2010mh,Azatov:2012bz} seems to be quite suitable to this aim: it assumes only that there are no new 
light states 
in the spectrum beyond a Higgs-like scalar (and thus the Higgs can decay only to pairs of SM particles) and that there is a (at least approximate) custodial symmetry.
A further request for the absence of (sizable) Higgs flavor-violating couplings is necessary to comply with the experimental data from FCNC processes.
If necessary, the first assumption can be relaxed by introducing in the effective Lagrangian any new light state which will be possibly discovered at the LHC.
Under the above hypotheses, the strength of the Higgs interactions is parametrized in terms of a set of parameters which must be determined experimentally.
In this enlarged parameter space, the SM corresponds to a point in the vicinity of which the theory stays perturbative up to very high energies.
The framework is sufficiently general to describe theories of composite Higgs, supersymmetric theories where the additional scalars are much heavier, and 
even models where the light scalar is not part of an $SU(2)_L$ doublet or is not related to EWSB like in the case of a dilaton.
From a practical point of view, all the Higgs production cross sections and decay fractions can be easily derived from a simple rescaling of  SM known
expressions. For convenience, we collect in the Appendix the relevant formulas for the simplified case in which only the overall strength of the Higgs coupling 
to vector bosons, $a = g_{hVV}/g_{hVV}^{SM}$, and to fermions, $c =g_{h\psi\psi}/g_{h\psi\psi}^{SM}$, are free to vary.

In this paper we have made a first attempt to estimate the precision that  the LHC can reach on $a$ and $c$ with 2012 data
by focusing on the $h\to \gamma\gamma$ channel, which is the most sensitive for a light Higgs. We have set the Higgs mass to the benchmark
value $m_h =120\,$GeV, and discussed how the results change in the range $120-130\,$GeV.
 Our first important conclusion is that exclusive analyses are much more powerful than inclusive ones both to put limits on and to precisely
measure the Higgs couplings. This is especially true for theories of non-standard Higgses (like fermiophobic models) where the importance of the VBF 
and associated production is enhanced compared to gluon fusion. A milder though significant improvement is however achieved even for a SM Higgs.
It is thus clear that performing analyses in ways that are as exclusive as possible is a crucial strategy for a precise determination of the Higgs couplings.
By injecting a SM signal in our simulation, we also found that $\gamma\gamma$ alone selects a second degenerate solution in the plane $(a,c)$.
The degeneracy can be broken only by adding additional channels to the fit. We find that $ZZ\to 4l$ seems to be the most powerful channel to this aim,
since it can lead to a precise determination of the coupling $a$. Other channels such as $WW$ and $\tau\tau$ turn out to be less precise and their inclusion
does not have a strong impact on the fit. Our extrapolation of the current analyses to $L= 40\,\text{fb}^{-1}$ (the total amount of integrated 
luminosity which might be obtained  by the end of 2012 by CMS and ATLAS together) shows that the second solution will be disfavored but is not eliminated.
Focusing on the solution centered at the SM point, we estimate that for $m_h =120\,$GeV the precision with which the couplings $a$ and $c$ can be determined 
at 68\% of probability is respectively $\sim 25\%$ and $\sim 100\%$.
More refined strategies or a larger amount of luminosity seem to be required in order to completely exclude the second solution and obtain a more precise determination
of $a$ and $c$.

\section*{Acknowledgments}

We would like to thank Emanuele Di Marco and Simone Gennai for important discussions and suggestions, and Christophe Grojean for discussions
and  useful comments on the manuscript.
The work of R.C. was partly supported by the ERC Advanced Grant No.~267985 \textit{Electroweak Symmetry Breaking, Flavour and Dark Matter: One Solution for 
Three Mysteries (DaMeSyFla)}.

\appendix
\section{Appendix}

We collect here the formulas needed to compute the Higgs production cross section and branching fractions in terms of  SM values for the case in which the overall strength of the Higgs coupling 
to vector bosons, $a = g_{hVV}/g_{hVV}^{SM}$, and to fermions, $c =g_{h\psi\psi}/g_{h\psi\psi}^{SM}$, are free to vary. This is a special simplified
scenario of the general parametrization of the Higgs couplings introduced in~\cite{Contino:2010mh,Azatov:2012bz}.
There are no new production modes nor new decay channels in addition to those present in the SM.

The expression of the four Higgs production cross section is given by a simple rescaling of the SM ones ($V = W,Z$):
\begin{equation}
\begin{split}
\sigma(gg\to h) &= c^2 \, \sigma(gg\to h)_{SM} \\[0.15cm]
\sigma(qq\to qq h) &= a^2 \, \sigma(qq\to qq h)_{SM} \\[0.15cm]
\sigma(q\bar q\to Vh) &= a^2 \, \sigma(q\bar q\to Vh)_{SM} \\[0.15cm]
\sigma(gg, q\bar q\to t\bar t h) &= c^2 \,\sigma(gg, q\bar q\to t\bar t h)_{SM} 
\end{split}
\end{equation}
The decay branching ratios are determined by a simple rescaling of the Higgs partial widths. The formulas for these latter are ($f$ denotes any of the
quarks and leptons of the SM):
\begin{equation}
\begin{split}
\Gamma(h\to VV) & = a^2 \, \Gamma(h\to VV)_{SM} \\[0.15cm]
\Gamma(h\to f\bar f) & = c^2 \, \Gamma(h\to  f\bar f)_{SM} \\[0.15cm]
\Gamma(h\to gg) & = c^2 \, \Gamma(h\to  gg)_{SM} \\[0.15cm]
\Gamma(h\to \gamma\gamma) & = \frac{\left|c\, A_f(m_h) + a\, A_W(m_h)\right|^2}{\left| A_f(m_h) + A_W(m_h)\right|^2} 
\, \Gamma(h\to \gamma\gamma)_{SM} \\[0.15cm]
\Gamma(h\to Z\gamma) & = \frac{\left|c\, B_f(m_h) + a\, B_W(m_h)\right|^2}{\left| B_f(m_h) + B_W(m_h)\right|^2} \, \Gamma(h\to Z\gamma) _{SM}
\end{split}
\end{equation}
so that $\Gamma_{tot}(h)$ is the sum of the above partial widths and $BR(h \to X) = \Gamma(h\to X)/\Gamma_{tot}(h)$.
The functions $A$ and $B$ are given at one loop by
\begin{align}
A_f(m_h) =& \, -\frac{8}{3}  \frac{4 m_t^2}{m_h^2} \left[ 1+ \left(1 - \frac{4 m_t^2}{m_h^2} \right) \times  f\left( \frac{4 m_t^2}{m_h^2} \right) \right] , \\[0.4cm]
A_W(m_h) =&\, 2+ 3  \times \frac{4 m_W^2}{m_h^2} \left[ 1+ \left(2 - \frac{4 m_W^2}{m_h^2} \right) \times f \left( \frac{4 m_W^2}{m_h^2} \right) \right], \\[0.4cm]
B_f (m_h) =&\,  - \frac{ 4 \left(\frac{1}{2} -  \frac{4}{3} \sin^2 \theta_W\right)}{\sin \theta_W \cos \theta_W}
\left[ I_1  \left( \frac{4 m_t^2}{m_h^2}, \frac{4 m_t^2}{m_Z^2} \right) -  I_2  \left( \frac{4 m_t^2}{m_h^2}, \frac{4 m_t^2}{m_Z^2} \right)  \right], \\[0.4cm]
\begin{split}
B_W(m_h) =&\, -\frac{\cos \theta_W}{\sin\theta_W}  \times \left\{  
\left(12 - 4 \tan^2 \theta_W\right) \times I_2 \left(\frac{4 m_W^2}{m_h^2}, \frac{4 m_W^2}{m_Z^2} \right)\right. \\[0.1cm]
 & + \left.  \left[ \left( 1 + \frac{2 m_h^2}{4 m_W^2} \right)  \tan^2 \theta_W  - \left(5+ \frac{2m_h^2}{4 m_W^2} \right) \right]   
    \times I_1 \left( \frac{4 m_W^2}{m_h^2}, \frac{4 m_W^2}{m_Z^2} \right) \right\}\, , 
\end{split}
\end{align}
where
\begin{align}
I_1(a,b) &= \frac{ab}{2(a-b)} + \frac{a^2 b^2}{2(a-b)^2} \left[f(a) -f(b)\right] + \frac{a^2 b}{(a-b)^2} \left[ g(a) -g(b) \right], \\[0.1cm]
I_2(a,b) &= -\frac{ab}{2 (a-b)} \left[ f(a) - f(b) \right] ,
\end{align}
with
\begin{equation}
f(x) = \begin{cases} 
\left[\sin^{-1} \left(1/\sqrt x \right) \right]^2 &{\rm for }  \ x \geq 1 \\[0.1cm]
-\frac{1}{4} \left[  \log \left( \frac{1+\sqrt{1-x}}{1-\sqrt{1-x}} \right) -i \pi \right]^2 &{\rm for }  \ x < 1,
\end{cases} 
\end{equation}
and
\begin{equation}
g(x) = \begin{cases} 
\sqrt{x-1} \sin^{-1} \left( 1/ \sqrt x  \right) &{\rm for }  \ x \geq 1 \\[0.1cm]
\frac{1}{2} \sqrt{1-x} \left[  \log \left( \frac{1+\sqrt{1-x}}{1-\sqrt{1-x}} \right) -i \pi \right]^2 &{\rm for }  \ x < 1.
\end{cases} 
\end{equation}
For a full discussion of these results, and expressions for more general cases where new fields can contribute to the loop functions, see for instance \cite{HiggsHunters}.

For simplicity,  we quote in Table~\ref{table:AB}  numerical values of the functions $A_{f,W}$ and $B_{f,W}$ for the mass range of interest.  Note that the contribution from gauge bosons in $h \to Z \gamma$ are on the order of 20 times larger than the contribution from fermions; in practice, modifications to this decay will have a negligible impact on results throughout the space explored in this analysis.
\begin{table}[htb]
\begin{center}
\begin{tabular}{c | c | c | c | c  }
$m_h$ (GeV)  & $A_f$ & $A_W$ & $B_f$ & $B_W$  \cr   \hline \hline
100 & -1.81 & 7.72 & 0.635 & -10.8   \\
110 & -1.82 & 7.93 & 0.638 & -11.2   \\
120 & -1.83 & 8.19 & 0.641 & -11.7   \\
130 & -1.84 & 8.53 & 0.644 & -12.3   \\
140 & -1.85 & 9.01 & 0.648 & -13.2   \\
150 & -1.86 & 9.76 & 0.652 & -14.7   \\
160 & -1.87 & 12.40 & 0.657 & -20.0   \\
\hline
\end{tabular}
\caption{\small
Numerical values for rescaling factors in loop-mediated processes $h \to \gamma \gamma$ and $h\to Z \gamma$.}
\label{table:AB}
\end{center}
\label{default}
\end{table}


\begin{thebibliography}{99}

\bibitem{compositeHiggs}
  D.~B.~Kaplan and H.~Georgi,
  Phys.\ Lett.\  B {\bf 136} (1984) 183.
S.~Dimopoulos and J.~Preskill,
  Nucl.\ Phys.\  B {\bf 199}, 206 (1982).
T.~Banks,
  Nucl.\ Phys.\  B {\bf 243}, 125 (1984).
D.~B.~Kaplan, H.~Georgi and S.~Dimopoulos,
  Phys.\ Lett.\  B {\bf 136}, 187 (1984).
H.~Georgi, D.~B.~Kaplan and P.~Galison,
  Phys.\ Lett.\  B {\bf 143}, 152 (1984).
H.~Georgi and D.~B.~Kaplan,
  Phys.\ Lett.\  B {\bf 145}, 216 (1984).
M.~J.~Dugan, H.~Georgi and D.~B.~Kaplan,
  Nucl.\ Phys.\  B {\bf 254}, 299 (1985).

\bibitem{Contino:2010mh}
  R.~Contino, C.~Grojean, M.~Moretti, F.~Piccinini and R.~Rattazzi,
  JHEP {\bf 1005} (2010) 089
  [arXiv:1002.1011 [hep-ph]].

\bibitem{Azatov:2012bz}
  A.~Azatov, R.~Contino and J.~Galloway,
  arXiv:1202.3415 [hep-ph].

\bibitem{previousCHL}
  J.~F.~Donoghue, C.~Ramirez and G.~Valencia,
  Phys.\ Rev.\ D {\bf 39} (1989) 1947;
  J.~F.~Donoghue and C.~Ramirez,
  Phys.\ Lett.\ B {\bf 234} (1990) 361;
  J.~Bagger, V.~D.~Barger, K.~-m.~Cheung, J.~F.~Gunion, T.~Han, G.~A.~Ladinsky, R.~Rosenfeld and C.~-P.~Yuan,
  Phys.\ Rev.\ D {\bf 52} (1995) 3878
  [hep-ph/9504426].


\bibitem{dilaton}
  E.~Halyo,
  Mod.\ Phys.\ Lett.\  A {\bf 8} (1993) 275.
  W.~D.~Goldberger, B.~Grinstein and W.~Skiba,
  Phys.\ Rev.\ Lett.\  {\bf 100} (2008) 111802
  [arXiv:0708.1463 [hep-ph]];
  L.~Vecchi,
  Phys.\ Rev.\ D {\bf 82} (2010) 076009
  [arXiv:1002.1721 [hep-ph]];
  B.~A.~Campbell, J.~Ellis and K.~A.~Olive,
  arXiv:1111.4495 [hep-ph].

\bibitem{Giudice:2007fh}
  G.~F.~Giudice, C.~Grojean, A.~Pomarol and R.~Rattazzi,
  JHEP {\bf 0706} (2007) 045
  [arXiv:hep-ph/0703164].



\bibitem{Zeppenfeld:2000td}
  D.~Zeppenfeld, R.~Kinnunen, A.~Nikitenko and E.~Richter-Was,
  Phys.\ Rev.\ D {\bf 62} (2000) 013009
  [hep-ph/0002036].

\bibitem{Conway:2002kk}
  J.~Conway {\it et al.}  [Precision Higgs Working Group of Snowmass 2001 Collaboration],
  eConf C {\bf 010630} (2001) P1WG2
  [hep-ph/0203206].

\bibitem{Belyaev:2002ua}
  A.~Belyaev and L.~Reina,
  JHEP {\bf 0208} (2002) 041
  [hep-ph/0205270].

\bibitem{duhrssen}
M. Duhrssen, 
``Prospects for the measurement of Higgs boson coupling parameters in the mass range from $110-190$\,GeV$/c^2$'',
ATL-PHYS-2003-030.

\bibitem{Duhrssen:2004cv}
  M.~Duhrssen, S.~Heinemeyer, H.~Logan, D.~Rainwater, G.~Weiglein and D.~Zeppenfeld,
  Phys.\ Rev.\  D {\bf 70} (2004) 113009
  [arXiv:hep-ph/0406323].

\bibitem{Lafaye:2009vr}
  R.~Lafaye, T.~Plehn, M.~Rauch, D.~Zerwas and M.~Duhrssen,
  JHEP {\bf 0908} (2009) 009
  [arXiv:0904.3866 [hep-ph]].

 \bibitem{Bock:2010nz}
  S.~Bock, R.~Lafaye, T.~Plehn, M.~Rauch, D.~Zerwas and P.~M.~Zerwas,
  Phys.\ Lett.\ B {\bf 694} (2010) 44
  [arXiv:1007.2645 [hep-ph]].



\bibitem{Carmi:2012yp}
  D.~Carmi, A.~Falkowski, E.~Kuflik and T.~Volansky,
  arXiv:1202.3144 [hep-ph].

\bibitem{Espinosa:2012ir}
  J.~R.~Espinosa, C.~Grojean, M.~Muhlleitner and M.~Trott,
  arXiv:1202.3697 [hep-ph].

\bibitem{Giardino:2012ww}
  P.~P.~Giardino, K.~Kannike, M.~Raidal and A.~Strumia,
  arXiv:1203.4254 [hep-ph].

\bibitem{Rauch:2012wa}
  M.~Rauch,
  arXiv:1203.6826 [hep-ph];
M. Klute, R. Lafaye, T. Plehn, M. Rauch, D. Zerwas, M. Duhrssen, in preparation.

\bibitem{Ellis:2012rx}
  J.~Ellis and T.~You,
  arXiv:1204.0464 [hep-ph].



\bibitem{Chatrchyan:2012tw} 
  S.~Chatrchyan {\it et al.}  [CMS Collaboration],
  ``Search for the standard model Higgs boson decaying into two photons in pp collisions at sqrt(s)=7 TeV,''
  arXiv:1202.1487 [hep-ex] (accepted by PLB).

\bibitem{CMS-PAS-HIG-12-002}
CMS Collaboration, ``Search for the fermiophobic model Higgs boson decaying into two photons," CMS-PAS HIG-12-002.


\bibitem{Alwall:2007st} 
  J.~Alwall, P.~Demin, S.~de Visscher, R.~Frederix, M.~Herquet, F.~Maltoni, T.~Plehn and D.~L.~Rainwater {\it et al.},
  JHEP {\bf 0709}, 028 (2007)
  [arXiv:0706.2334 [hep-ph]].

\bibitem{pythia} 
T.~Sj\"ostrand, S.~Mrenna, and P.Z.~Skands, PYTHIA 6.4 Physics and Manual, JHEP {\bf 0605}, 026 (2006) 
   {\tt doi:10.1088/1126-6708/2006/05/026}.

\bibitem{Alioli:2008tz} 
  S.~Alioli, P.~Nason, C.~Oleari and E.~Re,
  JHEP {\bf 0904}, 002 (2009)
  [arXiv:0812.0578 [hep-ph]].

\bibitem{Nason:2009ai} 
  P.~Nason and C.~Oleari,
  JHEP {\bf 1002}, 037 (2010)
  [arXiv:0911.5299 [hep-ph]].

\bibitem{CMS-PAS-HIG-11-030}
CMS Collaboration, ``Search for a Higgs boson decaying into two photons in the CMS detector'', CMS-PAS HIG-11-030.

\bibitem{D'Agostini:2003nk}
  G.~D'Agostini,
  ``Bayesian reasoning in data analysis: A critical introduction,''
  New Jersey, USA: World Scientific (2003) 329 p

\bibitem{LHCsummer}
ATLAS and CMS Collaborations and LHC Higgs Combination Group, \textit{Procedure for the LHC Higgs boson search combination in Summer 2011}, 
CMS-NOTE-2011/005; ATL-PHYS-PUB-2011-11 (2011).

\bibitem{CMS-PAS-HIG-11-021}
CMS Collaboration, ``Search for a Higgs boson decaying into two photons in the CMS detector'', CMS-PAS HIG-11-021.

\bibitem{Gabrielli:2012yz}
  E.~Gabrielli, B.~Mele and M.~Raidal,
  arXiv:1202.1796 [hep-ph].

\bibitem{Chatrchyan:2012ty}
  S.~Chatrchyan {\it et al.}  [CMS Collaboration],
  ``Search for the standard model Higgs boson decaying to a W pair in the fully leptonic final state in pp collisions at sqrt(s) = 7 TeV,''
  arXiv:1202.1489 [hep-ex] (accepted by PLB).

\bibitem{Chatrchyan:2012dg}
  S.~Chatrchyan {\it et al.}  [CMS Collaboration],
  ``Search for the standard model Higgs boson in the decay channel H to ZZ to 4 leptons in pp collisions at sqrt(s) = 7 TeV,''
  arXiv:1202.1997 [hep-ex] (accepted by PRL).


\bibitem{Chatrchyan:2012vp}
  S.~Chatrchyan {\it et al.}  [CMS Collaboration],
  ``Search for neutral Higgs bosons decaying to tau pairs in pp collisions at sqrt(s)=7 TeV,''
  arXiv:1202.4083 [hep-ex] (accepted by PLB).

\bibitem{Contino:2006qr}
  R.~Contino, L.~Da Rold and A.~Pomarol,
  Phys.\ Rev.\  D {\bf 75} (2007) 055014
  [arXiv:hep-ph/0612048].

\bibitem{CMS-PAS-HIG-12-008}
CMS Collaboration, ``Combined results of searches for a Higgs boson in the context of the standard model and beyond-standard models'', CMS-PAS HIG-12-008.

\bibitem{Lafaye:2004cn}
  R.~Lafaye, T.~Plehn and D.~Zerwas,
  hep-ph/0404282.

\bibitem{HiggsHunters} 
  J.~F.~Gunion, H.~E.~Haber, G.~L.~Kane and S.~Dawson,
  ``The Higgs Hunter's Guide,''
  Front.\ Phys.\  {\bf 80}, 1 (2000).

\end{thebibliography}
\end{document}